\documentclass[prd,amsfonts,nofootinbib,preprintnumbers,nobalancelastpage,superscriptaddress]{revtex4}
\usepackage{graphicx}
\usepackage{epsfig} 
\usepackage{xcolor}
\usepackage{rotating}
\usepackage{amsmath}
\usepackage{braket}
\usepackage{pdflscape}

\newcommand{\be}{\begin{equation}}
\newcommand{\ee}{\end{equation}}

\newcommand{\w}{\omega}

\newcommand{\Dmq}{\Delta m^2}
\newcommand{\eVq}{\ensuremath{\text{eV}^2}}

\begin{document}
\preprint{YITP-SB-15-38}
\title{Conditions for Statistical Determination of the Neutrino
Mass Spectrum in Radiative Emission of Neutrino Pairs in Atoms} 
\author{Ningqiang Song}
\email{ningqiang.song@stonybrook.edu}
\affiliation{%
  C.N.~Yang Institute for Theoretical Physics, SUNY at Stony Brook,
  Stony Brook, NY 11794-3840, USA}
\author{R. Boyero Garcia}\email{robertobg@usal.es}
\affiliation{Centro de L\'aseres Pulsados, CLPU, Parque Cient\'ifico, 37185 Villamayor, Salamanca, Spain.}
\author{J.\ J.\ Gomez-Cadenas}\email{gomez@mail.cern.ch,}
\affiliation{
Instituto de F\'{\i}sica Corpuscular (IFIC), CSIC \& Universitat de Valencia,
Calle Catedr\'atico Jos\'e Beltr\'an, 2, 46980 Paterna, Valencia, Spain}
\author{M.\ C.\ Gonzalez--Garcia} \email{maria.gonzalez-garcia@stonybrook.edu}
\affiliation{%
  Instituci\'o Catalana de Recerca i Estudis Avan\c{c}ats (ICREA),}
\affiliation {Departament d'Estructura i Constituents de la Mat\`eria, 
Universitat
  de Barcelona, 647 Diagonal, E-08028 Barcelona, Spain}
\affiliation{%
  C.N.~Yang Institute for Theoretical Physics, SUNY at Stony Brook,
  Stony Brook, NY 11794-3840, USA}
\author{A. Peralta Conde}
\email{aperalta@clpu.es}
\affiliation{Centro de L\'aseres Pulsados, CLPU, Parque Cient\'ifico, 37185 Villamayor, Salamanca, Spain.}
\author{Josep Taron}
\email{taron@ecm.ub.edu}
\affiliation{Departament d'Estructura i Constituents de la Mat\`eria, 
Universitat  de Barcelona, 647 Diagonal, E-08028 Barcelona, Spain}
\begin{abstract}
The photon spectrum in macrocoherent atomic de-excitation via
radiative emission of neutrino pairs (RENP) has been proposed as a
sensitive probe of the neutrino mass spectrum, capable of competing
with conventional neutrino experiments.  In this paper we revisit this
intriguing technique in order to quantify the requirements for
statistical determination of some of the properties of the neutrino
spectrum, in particular the neutrino mass scale and the mass
ordering. Our results are sobering. We find that, even under ideal
conditions, the determination of neutrino parameters needs
experimental live times of the order of days to years for several
laser frequencies, assuming a target of volume of order 100 cm$^3$
containing about $10^{21}$ atoms per cubic centimeter in a totally
coherent state with maximum value of the electric field in the
target. Such conditions seem to be, as of today, way beyond the reach
of our current technology.

\end{abstract}
\maketitle
\section{Introduction}

Neutrino oscillation experiments have now established beyond doubt
that neutrinos are massive and there is leptonic flavour violation in
their propagation ~\cite{Pontecorvo:1967fh,Gribov:1968kq} (see
~\cite{GonzalezGarcia:2007ib} for an overview).
A consistent description of the global data on neutrino
oscillations is possible by assuming that the three known
neutrinos ($\nu_e$, $\nu_\mu$, $\nu_\tau$) are linear
quantum superposition of three massive states $\nu_i$ ($i=1,2,3$)
with masses $m_i$. Consequently, a leptonic mixing matrix is present
in the weak charged current interactions~\cite{Maki:1962mu,
  Kobayashi:1973fv} of the mass eigenstates, 
which can be parametrized as~\cite{PDG}:
\begin{equation}
  \label{eq:matrix}
  U =
  \begin{pmatrix}
    c_{12} c_{13}
    & s_{12} c_{13}
    & s_{13} e^{-i\delta_\text{CP}}
    \\
    - s_{12} c_{23} - c_{12} s_{13} s_{23} e^{i\delta_\text{CP}}
    & \hphantom{+} c_{12} c_{23} - s_{12} s_{13} s_{23}
    e^{i\delta_\text{CP}}
    & c_{13} s_{23}
    \\
    \hphantom{+} s_{12} s_{23} - c_{12} s_{13} c_{23} e^{i\delta_\text{CP}}
    & - c_{12} s_{23} - s_{12} s_{13} c_{23} e^{i\delta_\text{CP}}
    & c_{13} c_{23}
  \end{pmatrix} \begin{pmatrix}1 & 0 & 0 \\
0  & e^{i\eta_1} & 0 \\ 
0  & 0 & e^{i\eta_2} \end{pmatrix}
\end{equation}
where $c_{ij} \equiv \cos\theta_{ij}$ and $s_{ij} \equiv
\sin\theta_{ij}$.  The phases $\eta_{i}$ are only non-zero if neutrinos
are Majorana particles.
If one chooses the convention where the angles 
$\theta_{ij}$ are taken to lie in the first quadrant, $\theta_{ij} \in [0, \pi/2]$,  and the CP phases $\delta_\text{CP},\eta_1,\eta_2 \in [0, 2\pi]$, 
then $\Delta m^2_{21}=m_2^2-m_1^2>0$ by convention, and  $\Delta m^2_{31}$
can be positive or negative.
It is customary to refer to the first option as Normal Ordering (NO), and
to the second one as Inverted Ordering (IO).

At present the global analysis of neutrino oscillation data yields
the three-sigma ranges for the parameters \cite{Gonzalez-Garcia:2014bfa}
\begin{equation}
    \begin{tabular}{l|c}
      \hline\hline
       & $3\sigma$ range
      \\
      \hline
      \rule{0pt}{4mm}\ignorespaces
      $\sin^2\theta_{12}$
      & $0.270 \to 0.344$
      \\[3mm]
      $\sin^2\theta_{23}$
         & $0.385 \to 0.644$
         \\[3mm]
      $\sin^2\theta_{13}$
         & $0.0188 \to 0.0251$
         \\[3mm]
      $\delta_\text{CP}/^\circ$
         & $\hphantom{00}0 \to 360$
      \\[3mm]
      $\dfrac{\Dmq_{21}}{10^{-5}~\eVq}$
      & $7.02 \to 8.09$
      \\[3mm]
      $\dfrac{\Dmq_{3\ell}}{10^{-3}~\eVq}$
      & $\begin{bmatrix}
        +2.325 \to +2.599\\[-2pt]
        -2.590 \to -2.307
      \end{bmatrix}$
      \\[3mm]
      \hline\hline
    \end{tabular}\; ,
\label{eq:oscpar} 
\end{equation}
but gives no information on the Majorana phases nor on the Dirac or Majorana
nature of the neutrino. They do not provide a measurement of the 
absolute neutrino masses as well, but only of their differences.
In the table $\Dmq_{3\ell}$ corresponds to the largest mass splitting
(in absolute value) with $\ell=1$ for NO and $\ell=2$ for IO.
As seen from  the table, at present, oscillation experiments have not
provided us with information  of the ordering either.

The determination of the ordering and the 
CP violating phase $\delta_\text{CP}$ is the main goal of ongoing 
long baseline  (LBL) oscillation experiments \cite{Adamson:2014vgd,Abe:2015awa,Patterson:2012zs}   which are sensitive to those in some part of the
parameter space. Definite knowledge  is better guaranteed in future 
projects \cite{Abe:2015zbg,Bass:2013vcg}. 

Concerning the determination of the absolute mass scale 
in laboratory experiments, the standard approach is the search for
the distortion of the end point of the electron spectrum in tritium beta decay.
At present the most precise experiments~\cite{Bonn:2001tw,Lobashev:2001uu} 
have given no indication in favor of distortion what sets an
upper limit 
\begin{equation}
    \label{eq:nuelim}
    m_{\nu_e}=\left[\sum_i m^2_i |U_{ei}|^2\right]^{1/2} <2.2~\text{eV} \; . 
\end{equation}
The ongoing KATRIN experiment \cite{Osipowicz:2001sq}, is expected to achieve 
an estimated sensitivity limit: $m_\beta \sim 0.3$ eV.

The most precise probe of the nature of the neutrino is 
the search of neutrino-less double beta decay for verification of 
lepton number violation which is related to neutrino Majorana masses
(for a recent review see Ref.~\cite{GomezCadenas:2011it}).
So far this decay has not been observed and the strongest
bounds arise from experiments using $^{76}$Ge~\cite{Macolino:2013ifa}, 
$^{136}$Xe \cite{Gando:2012zm,Albert:2013gpz}, 
and $^{130}$Te\cite{Alfonso:2015wka}. 
For the case in which the only effective lepton number violation at
low energies is induced by the Majorana mass term for the neutrinos, 
the rate of $0\nu\beta\beta$ decay is proportional to the
\emph{effective Majorana mass of $\nu_e$}, and the experimental
bounds on the corresponding lifetimes can be translated in constraints
on the combination  
\begin{equation}
    m_{ee} 
= \left| \sum_i m_i U_{ei}^2 \right|\lesssim 0.14\to 0.76 \; \text{eV}\;, 
\end{equation}
which, in addition to the masses and mixing parameters that affect the
tritium beta decay spectrum, depends also on two combinations of the  
CP violating phases $\delta_\text{CP}$ and $\eta_i$. 

An unexpected new way to explore fundamental neutrino physics may come from the field of quantum optics, thanks to recent technological advances. The key concept behind the intriguing possibility is the small energy difference between the levels 
in the atom or molecule, which allows for large relative effects associated with
the small neutrino masses in the energy released in level transitions. This, in turn, opens
 up the possibility of precision neutrino mass spectroscopy, as
proposed by
Ref.~\cite{Yoshimura:2011ri,Fukumi:2012rn,Dinh:2012qb}.

The relevant process in this case is the atomic de-excitation via
radiative emission of neutrino pairs (RENP): $|e \rangle \rightarrow
|g\rangle + \gamma + \nu_i\bar \nu_j$.  The rate of this process can
be made measurable if macro-coherence of the atomic target can be
achieved \cite{Yoshimura:2012tm,Fukumi:2012rn}. The proposal is to
reach such macro-coherent emission of radiative neutrino pairs via
stimulation by irradiation of two trigger lasers of frequencies
$\omega, \omega'$ constrained by $\omega + \omega' =
\epsilon_{eg}/\hbar \,, \omega < \omega'$, with $E_{eg} = E_e - E_g$
being the energy difference of initial and final levels. With this
set-up the energy of the emitted photon in the de-excitation is given
by the smaller laser frequency $\omega$ and therefore it 
can be very precisely known.
Furthermore, neglecting atomic recoil, energy-momentum conservation  
implies that  each time the energy of the emitted photon  decreases 
below $\omega_{ij}$ with
\begin{equation}
\omega_{ij} = \frac{E_{eg}}{2} - \frac{(m_i +m_j)^2}{2E_{eg}}
\end{equation}
a new channel (this is, emission of another pair of massive neutrino spices) 
is kinematically open. 

Location of these threshold energies, by changing the laser frequency 
is, in principle possible, since the laser frequency, and therefore the emitted 
photon energy,  is known to high precision. Consequently once the six
$\omega_{ij}$ are measured, the spectrum of the neutrino masses could be fully
identified. It has been argued that this method is ultimately
capable of determining the neutrino mass scale, the mass ordering, 
the Dirac vs Majorana nature, as well as of  measuring the Majorana  
CP violating  phases \cite{Yoshimura:2011ri,Fukumi:2012rn,Dinh:2012qb}.

In this article we revisit this proposal with the aim at quantifying
the requirements for statistical determination of some of these
properties of the neutrino spectrum, in particular the neutrino mass
scale and the mass ordering. To do so we will review in
Sec.~\ref{sec:derivation} the derivation of the rate for RENP and the
corresponding photon energy spectrum.  Section \ref{sec:results}
contains our quantitative results and conclusions.

\section{Photon Energy Rate in RENP  and 
Neutrino Spectrum}
\label{sec:derivation}
The expected rate for RENP and the energy spectrum of the emitted photon
has been derived in Refs.~\cite{Fukumi:2012rn,Dinh:2012qb}  
and we have reproduced it (up to an overall factor 4). 
For the sake of completeness we summarize here the main elements 
and assumptions entering the derivation. 

The starting point is the effective Hamiltonian describing the atomic
transition $|e \rangle \rightarrow |g\rangle + \gamma + \nu_i\bar\nu_j$ 
assuming that the process cannot proceed directly but only via 
an intermediate {\sl virtual} state $|p\rangle$  with $E_p>E_e>E_g$, and
that the transition betaween  $|p\rangle$ and  $|g\rangle$ is 
of type E1 and leads to  the emission of the photon while 
the transition between  $|e\rangle$ and $|p\rangle$ is of type M1
leading to the emission of the neutrino pair. In this case, after integrating
out the intermediate state $|p\rangle$ in the Markovian 
and slow varying envelope approximation (see appendix A in 
\cite{Yoshimura:2012tm}), the
Schroedinger equation for the effective two--level atomic system state, 
$|\psi(x,t)\rangle
=c_e(x,t)|e\rangle+c_g(x,t)|e\rangle$, 
\begin{equation}
\frac{d}{dt}\psi(x,t)\equiv 
\frac{d}{dt}\left(\begin{array}{c} c_e(x,t)\\c_g(x,t)\end{array} \right)
=-i \,H_{\rm RENP}(x,t) 
\,
\left(\begin{array}{c} c_e(x,t)\\c_g(x,t)\end{array} \right)\; , 
\end{equation}
where $H_{\rm RENP}(x,t)$ takes the matrix form   
\begin{equation}
H_{\rm RENP}(x,t)=H^{\rm R}_{eg}(x,t)
\frac{\sigma_1-i\sigma_{2}}{2}  \; , 
\label{eq:renph}
\end{equation}
here $\sigma_i$ are the Pauli matrices, and 
\begin{equation}
H^{\rm R}_{eg}
=-\frac{G_F}{\sqrt{2}}{ \vec d_{gp}\cdot\vec{\widetilde{E}}^*(x,t)} 
\frac{1}{E+E'+E_{pe}}\,[\bar u^\lambda_i(p)\gamma_\mu (1-\gamma_5) v^{\lambda'}_j(p')]\,
(v_{ij} J^\mu_{V,pe}-a_{ij}J^\mu_{A,pe}) \, \exp^{i(\w+E+E'-E_{eg})t}
\exp^{-i(\vec p+\vec p'+\vec k)\vec x}\; .
\end{equation}
$\vec{\widetilde{E}}^*(x,t)$ is the amplitude of the electric field,
while $(\w,\vec k)$ is the four momentum of the photon.
Implicit in this expression is the hypothesis that the RENP transition is
driven by two lasers, one of which must have the frequency and wave number 
of the  emitted photon (more below). 
$E_{ab}=E_a-E_b$ is the energy difference between  two of the 
atomic levels, and 
\begin{eqnarray}
&&
\langle g| \vec d|p\rangle=\vec d_{gp}\;, 
\;\;\;\;\;\; \;\;\;\;\;\; \;\;\;\;\;\; \;\;\;\;\;\; \;\;\;\;\;\;
\;\;\;\;\;\; \;\;\;\;
\langle p |\bar f_e({ x})\gamma^\mu (\gamma^5) f_e(x')|e\rangle =
\delta^3(x-x')J^\mu_{V(A),pe}  
\\
&&v_{ij}=U^*_{ei}U_{ej}-\delta_{ij}(\frac{1}{2}-2 \sin^2\theta_w) \,
\;\;\;\;\;\;  \;\;\;\;\;\; a_{ij}=U^*_{ei}U_{ej}-\frac{1}{2}\delta_{ij} \; . 
\end{eqnarray}
$\vec d$ is the electric dipole moment operator, and 
$f_e$ is the electron field. In defining the 
electron atomic currents, $J^\mu_{V(A)}$, we have implicitly assumed that the 
spatial atomic wave function is  concentrated around the atomic position 
$\vec x$  so we have approximated it as a delta function. 
In the non-relativistic limit for the electron field it 
can be shown that $J^\mu_V=0=J^0_A$  while  $\vec J_{A,pe}=
\langle p|2\vec S |e\rangle $, where $\vec S$ is 
the  spin operator.

For a single atom at position $\vec x_a$  at time $t$
the transition amplitude from an initial atomic state of wave function 
$\psi^a_f(x_a)$  to a final atomic state $\psi^a_i(x_a)$ at first
order in perturbation theory is 
\begin{eqnarray}
A^a&=&\int_{-\infty}^\infty  H^{\rm R}_{eg}(x_a,t') dt'\nonumber \\
&\simeq&  - \frac{G_F}{\sqrt{2}} { \vec d_{gp}\cdot\vec{\tilde{E}}^*(x_a,t) } 
\frac{1}{\w-E_{pg}}\,[\bar u^\lambda_i(p)\gamma_\mu (1-\gamma_5) v^{\lambda'}_j(p')]\,a_{ij}\,J^\mu_{A,pe}\, \,
\left[(\psi^a_f(x_a))^\dagger \frac{\sigma_1-i \sigma_2}{2}\psi^a_i(x_a)\right] 
\nonumber \\
&&
\times \, \exp^{-i(\vec p+\vec p'+\vec k)\vec x_a}\,
(2\pi)\,\delta(E+E'+\w-E_{eg})
\; . 
\end{eqnarray}
where the energy momentum conservation condition implies $E+E'+E_{pe}={ E_{pg}-\w}$,
and it is assumed that the time scale for the 
transition is much shorter than the characteristic time variation of 
the electric field amplitude. We have introduced the atomic 
Bloch vector $\vec r^a(x_a,t)$ as: 
\begin{equation}
\left[(\psi^a_f(x_a))^\dagger \frac{\sigma_1-i \sigma_2}{2}\psi^a_i(x_a)\right]={ c^a_e(x_a,t_i) [c^a_g(x_a,t_f)]^*}\equiv \frac{r_1^a(x_a,t)-i r_2^a(x_a,t)}{2}
\end{equation}

The expression above is valid for emission from a single atom. For an 
ensemble of atoms in a volume $V$ centered in $\vec x$, the amplitude 
is the superposition of the contribution  of the $N$ atoms in the volume.
Following  Ref.~\cite{Fukumi:2012rn} 
one can approximate the summation as
${\displaystyle \sum_a}\exp^{-i(\vec p+\vec p'+\vec k)\vec x_a}
\simeq \frac{N}{V}\int dV \exp^{-i(\vec p+\vec p'+\vec k)\vec x_a}
\rightarrow N/V (2\pi)^3\delta(\vec p+\vec p'+\vec k)$. In this limit
\begin{equation}
\sum_a A^a= {\cal M}(x,t) 
(2\pi)^4 
\delta(E+E'+\w-E_{eg})\delta(\vec p+\vec p'+\vec k)
\end{equation}
where 
\begin{equation}
{\cal M}(x,t)=-\frac{G_F}{\sqrt{2}} { \vec d_{gp}\cdot\vec{\tilde{E}}^*(x,t)}  
\frac{1}{{ \w-E_{pg}}}\,[\bar u_i(p)\gamma_\mu (1-\gamma_5) v_j(p')]\,
a_{ij}{ J^\mu_{A,pe}} \frac{R_1(x)-iR_2(x)}{2}
\end{equation}
with the definition 
\begin{equation}
\sum_a \left[r_1^a(x_a,t)-i r_2^a(x_a,t)\right]\exp^{-i(\vec p+\vec p'+\vec k)\vec x_a}\equiv\left[R_1(x,t)-iR_2(x,t)\right] 
(2\pi)^3\delta(\vec p+\vec p'+\vec k)
\equiv  n(x) 
\left[r_1(x,t)-i r_1(x,t)\right]\; , 
\end{equation}
where $\vec R$ is the vector characterizing the medium 
``polarization'', $n(x)=N/V$ is the local density of the medium, 
so  $\vec r(x,t)$ is the mean value of $\vec R$  per atom.

As mentioned above the set-up to stimulate RENP is to radiate the
atomic medium (the target) with two  counter-propagating trigger lasers 
of frequencies $\w_1$ and $\w_2$ which verify $\w_1+\w_2=E_{eg}$, so the 
emitted photon has {$\w=\w_1$} and it is emitted in the direction of laser, 
$\vec k=\vec k_1$, with $|\vec k_1|=\w_1$.
Furthermore energy-momentum conservation implies $E+E'=\w_2$ and
$\vec k_1=-(\vec p +\vec p')$ , thus consequently for massive neutrinos
$\w_1<\w_2$. 

The number of stimulated transitions (ie the number of
single photons of frequency $\w$ 
emitted recoiling against the undetected neutrinos) per unit time and 
unit volume is
\begin{equation}
\frac{dN_\gamma(\omega)}{dt d^3x}=
\frac{1}{2J_e+1}\sum_{m_e}\sum_{m_p}\sum_{m_g}\sum_{\lambda,\lambda'} \int |{\cal M}|^2 
{\frac{d^3 p}{(2\pi)^3 2 E} \frac{d^3 p'}{(2\pi)^3 2 E'}} (2\pi)^4 
\delta^3(\vec p+\vec p'+\vec k)\delta(E+E'+\w-E_{eg}) \;
\end{equation}
where we denote by $m_{e,p,g}$ the third component of the angular momentum
of the electron in the corresponding atomic states, and we have averaged
over the initial angular configurations ($2J_e +1$) 
and summed over final ones. We have also 
summed over all possible configurations in the intermediate state $|p\rangle$.
Assuming isotropy one introduces the atomic spin factor $C_{ep}$ as, 
\begin{equation} 
{\sum_{m_p} \sum_{m_e} 
J^\mu_{A,pe} 
(J^\nu_{A,pe})^\dagger=\sum_{m_p} \sum_{m_e} 
4\langle p | S^i |e \rangle \langle e |S^j| p\rangle \equiv 
\frac{4}{3} \delta^{ij} (2J_e+1) (2J_p+1) C_{ep} \;.}
\end{equation}
Altogether 
\begin{eqnarray}
\frac{dN_\gamma(\omega)}{dt}&=&\frac{2 G_F^2}{\pi}\, (2 J_p+1)\, C_{ep}\,  
\int d^3x\, \left|\vec d_{pg}\cdot\vec{\widetilde{E}}(x,t)\right|^2 
\left|\frac{R_1(x,t)-iR_2(x,t)}{2}\right|^2 \, I(\w)\, \nonumber \\
&=&
6\, G_F^2\, V_{\rm tar}\, n^3 \,(2 J_p+1)\, C_{ep} \, \gamma_{pg}\, 
\frac{E_{eg}}{E_{pg}^3} \,I(\w) \, { \eta_\omega(t)}\, \nonumber \\ 
&=&\, 
0.464\, {\rm s}^{-1} \; (2J_p+1)\, C_{ep}\,
\left(\frac{V_{\rm tar}}{10^{2}\,{\rm cm}^{3}}\right) 
\left(\frac{n}{10^{21}\,{\rm cm}^{-3}}\right)^3 
\left(\frac{\gamma_{pg}} {10^8\,{\rm s}^{-1}}\right) 
\left(\frac{E_{eg}}{\rm eV}\right)
\left(\frac{\rm eV}{E_{pg}}\right)^3  \,I(\w) \,{ \eta_\w(t)}
\; . 
\end{eqnarray}
 In the second equality we have introduced the dimensionless factor 
\begin{equation}
\eta_\omega(t)=
\frac{1}{V_{\rm tar}}\int  d^3x  \frac{|r_1(x,t)|^2+|r_2(x,t)|^2}{4}
\frac{|\vec{\widetilde{E}}(x,t)|^2}{nE_{eg}}
\simeq
\frac{1}{L}\int_0^L dx  \frac{|r_1(x,t)|^2+|r_2(x,t)|^2}{4}
\frac{|\vec{\widetilde{E}}(x,t)|^2}{nE_{eg}}
\label{eq:renpfinal}
\end{equation}
where the second equality holds for a  long thin cylindrical target of 
total volume $V_{\rm tar}$.  
$\eta_\w(t)$  quantifies how many of the atoms in the target are coherently 
set in a state characterized by the same value of $r_i$ and how much
the energy density of the electric field in the medium, which is
$\propto |\vec E(x,t)|^2$,   
approaches the maximum value $E_{eg}\,n$.  
Both $\vec R$ and $\vec {\widetilde{E}}(x,t)$ have to be obtained independently by solving the coupled Bloch-Maxwell equations for the electromagnetic field in the 
presence of the atomic medium 
polarization (see Ref.~\cite{Fukumi:2012rn} 
and references therein). Furthermore we have introduced the spontaneous 
dipole transition rate $\gamma_{pg}= E^3_{pg}|\vec d_{pg}|^2/(3\pi)$ which 
is  experimentally measurable. 
\vskip 0.2cm
$I(\w)$ is the spectrum function, which (in agreement with 
Ref.~\cite{Dinh:2012qb}) reads
\begin{eqnarray}
I(\omega)&=&\frac{1}{(\w-E_{eg})^2}\sum_{ij} \Delta_{ij}(\w)
\left[|a_{ij}|^2 I_{ij}(\w)-\delta_M m_im_j {\rm Re}(a_{ij})^2\right] \;
\Theta\left({\omega}-\frac{E_{eg}}{2}+\frac{(m_i+m_j)^2}{2E_{eg}}\right)
\\
\Delta_{ij}(\w)&=&
\frac{\left[\left(E_{eg}(E_{eg}-2 \w)-(m_i+m_j)^2\right)
\left(E_{eg}(E_{eg}-2 \w)-(m_i-m_j)^2\right)
\right]^{1/2}}{E_{eg}(E_{eg}-2\w)} 
\\
I_{ij}(\w)&=&
\frac{1}{3}\left[E_{eg}(E_{eg}-2\w)+\frac{1}{2}\w^2
-\frac{1}{6}\w^2\Delta^2_{ij}(\w)-\frac{1}{2}(m_i^2+m_j^2)
-\frac{1}{2}\frac{(E_{eg}-\w)^2}{E^2_{eg}(E_{eg}-2\w)^2}(m_i^2-m_j^2)^2  
\right]\; , 
\end{eqnarray}
\begin{equation}
N_\gamma(\omega)= 0.464\, {\rm s}^{-1} \frac{T}{\rm s} \; (2J_p+1)\, C_{ep}\,
\left(\frac{V_{\rm tar}}{10^{2}\,{\rm cm}^{3}}\right) 
\left(\frac{n}{10^{21}\,{\rm cm}^{-3}}\right)^3 
\left(\frac{\gamma_{pg}} {10^8\,{\rm s}^{-1}}\right) 
\left(\frac{E_{eg}}{\rm eV}\right)
\left(\frac{\rm eV}{E_{pg}}\right)^3  \,I(\w) \, 
\langle \eta_\w \rangle 
\; ,
\end{equation}  
where we denote by $\langle \eta_\w \rangle$ the time average of 
$\eta_\w(t)$ along the duration of the laser irradiation. 

The requirements of the type of atomic transitions for RENP
imposes important constraints on the possible target atoms.
Two possible atomic candidates have been identified in the literature:
Yb and Xe, for which atomic levels with the required quantum numbers 
exist \cite{nist}:
\begin{equation}
\begin{array}{|c||c|c|c|c||c|c|c|c|}
\hline
 & \multicolumn{4}{c||} 
{\rm Xe}  & \multicolumn{4}{c|} {\rm Yb} \\
\hline 
& {\rm Config} &{\rm Term} & J & {\rm Level}({\rm cm}^{-1})&
{\rm Config} &{\rm Term} & J & {\rm Level}({\rm cm}^{-1})\\
\hline
|g\rangle & 5p^6 & ^1S & 0 & 0.0000 &
4f^{14}(^1S)6s^2 & ^1S & 0 & 0.0000 \\
|e\rangle & 5p^5(^2P_{3/2})6s & ^2[3/2]^o & 2 & 67067.547&
4f^{14}(^1S)6s6p & ^3P^o & 0 & 17288.439 \\
|p\rangle & 5p^5(^2P_{3/2})6s & ^2[3/2]^o & 1 & 68045.156&
4f^{14}(^1S)6s6p & ^3P^o & 1 &  17992.007 \\
\hline
E_{eg}({\rm eV})
& \multicolumn{4}{c||} {8.31632} 
& \multicolumn{4}{c|} {2.14349} 
\\
E_{pg}({\rm eV})
& \multicolumn{4}{c||} {8.43653} 
& \multicolumn{4}{c|} {2.23072} 
\\
\gamma_{pg} (10^{8} {\rm s}^{-1})
& \multicolumn{4}{c||} {2.73} 
& \multicolumn{4}{c|} {0.0115} 
\\
(2J_p+1) C_{ep}
& \multicolumn{4}{c||} 2 
& \multicolumn{4}{c|} 2 
\\
\hline 
\end{array}
\end{equation}

\begin{figure}
\centering
\includegraphics[width=0.5\textwidth]{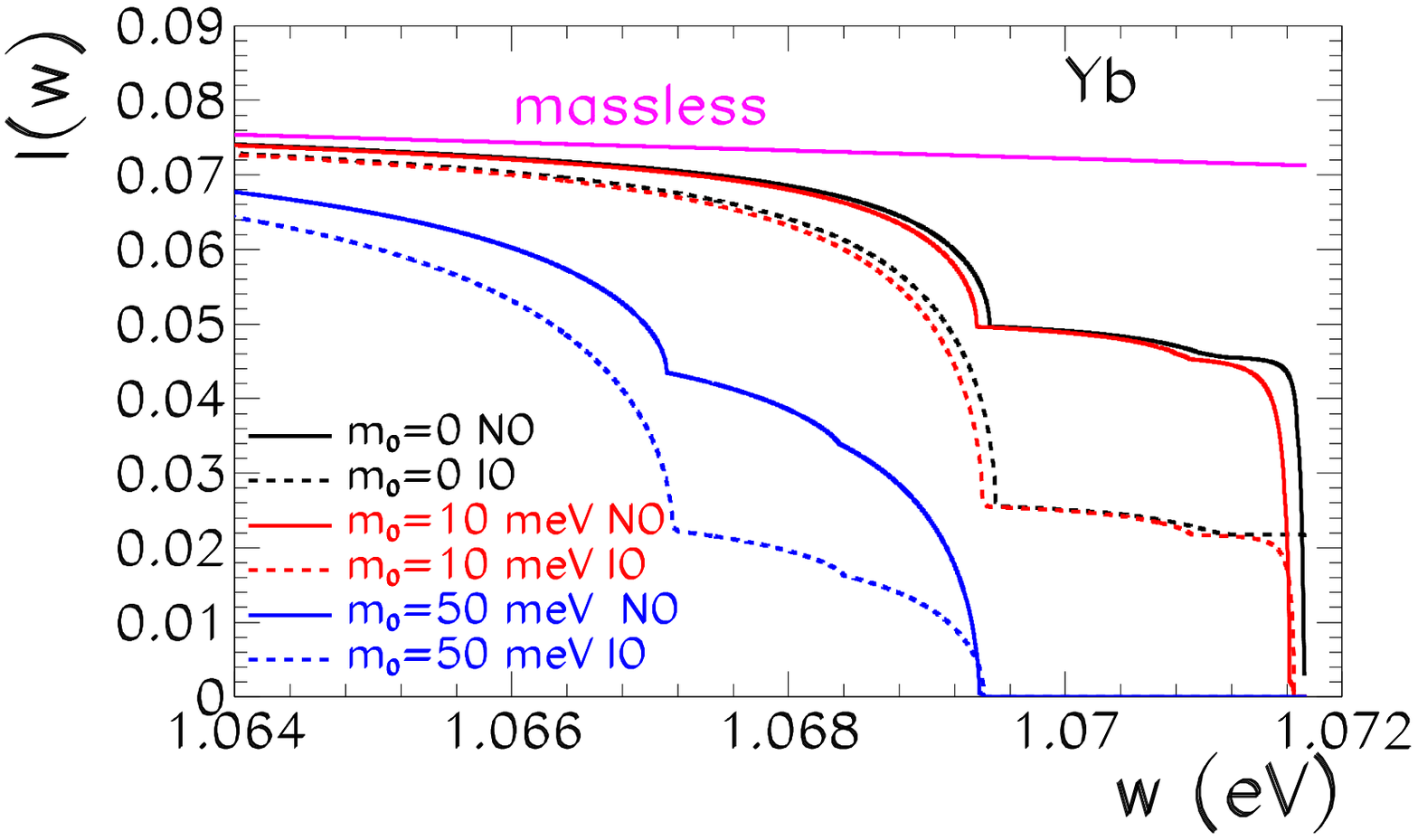}\vglue -0.3cm
\includegraphics[width=0.5\textwidth]{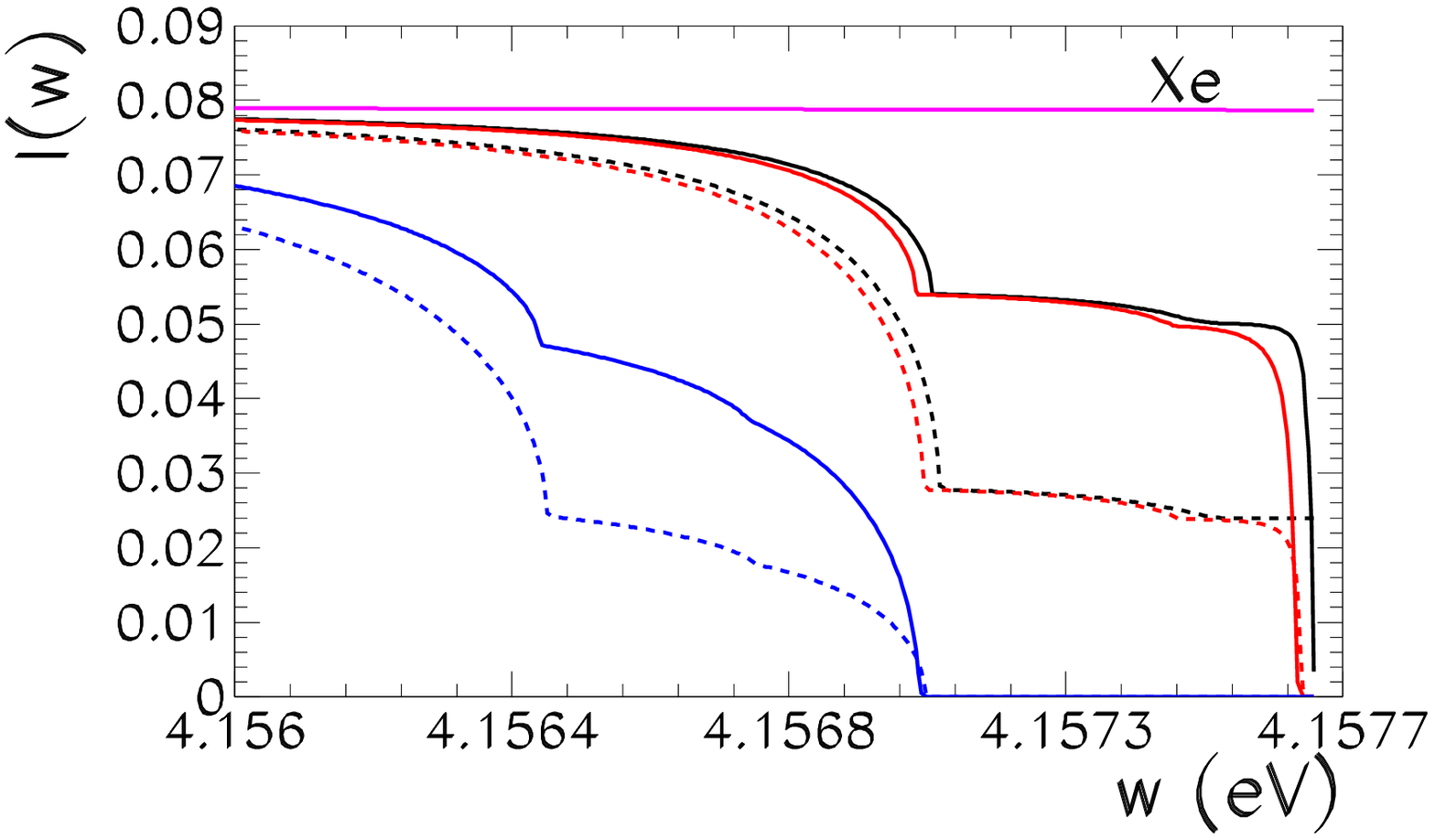}
\caption{RENP Spectral function $I(\w)$ for Yb (upper panel) and
Xe (lower panel) for different values of the lightest neutrino mass
$m_0$ and for both orderings, as labeled in the figure.
The curves correspond to the best fit oscillation parameters as
given in Eq.~(\ref{eq:oscpar}) and to Dirac neutrinos. The corresponding
curves for Majorana neutrinos are practically indistinguishable.
For illustration we also show the spectrum for three massless neutrinos.} 
\label{fig:spec}
\end{figure}

We plot in Fig~\ref{fig:spec} the RENP spectral function 
$I(\w)$ for these two nuclei near the end point for three different
values of the lightest neutrino mass, $m_0$, and for the best fit values
of the oscillation parameters in Eq.~(\ref{eq:oscpar}) for both orderings.
The spectrum shows the clear dependence of the end-point frequency 
on  $m_0$ as well as the differences between  NO and IO which mainly results 
in  different normalization for both spectra. 
The curves in the figure correspond to Dirac neutrinos, but the corresponding
curves for Majorana neutrinos are practically indistinguishable for those 
in the figure. 

This figure illustrates the potential of RENP to determine the neutrino
mass spectrum as well as the main differences between the two nuclei.
First because of the larger value of $E_{eg}$ the resolution 
in $\omega$ (the frequency of the trigger laser) required to resolve the 
threshold positions must be better for 
Xe than for Yb. On the other hand, because of the 
larger decay 
rate $\gamma_{pg}$ the expected RENP event rate is larger for Xe.

As seen in Eq.~(\ref{eq:renpfinal}) the RENP event rate grows as the 
third power of  the number density of atoms in the target, provided 
that both the 
amplitude of the electric field in the target acquires a value close to 
the maximum allowed,  and that the medium atomic polarization approaches 
its macro-coherent value.  In what follows we will quantify the 
final requirement on this product of factors to statistically determine 
the neutrino mass scale $m_0$ and the ordering.

\section{Results}
\label{sec:results}
\subsection{Determination of the Neutrino Mass Scale}
We  start by building the simplest observable sensitive to the
neutrino mass scale, this is to the value of the end-point frequency. 

In order to locate the end-point frequency of the RENP 
spectrum  we foresee a naive  experiment starting at a trigger frequency 
corresponding to  the end-point frequency for $m_0=0$. 
Clearly no RENP event should be observed  at such frequency.  
One then repeats the experiment
lowering one of the laser frequencies (while increasing the other keeping
the condition $\omega_1+\omega_2=E_{eg}$) 
in intervals of $\Delta_\omega$ until an  observation occurs. 
If we call $\omega_+$  to the maximum frequency for which 
no event is observed and $\omega_{-}=\omega_+-\Delta_\omega$  
the highest frequency for which some RENP events are observed, the CL at which 
this naive experiment can determine the neutrino mass scale $m_0$ with 
resolution $\pm\sigma_{m0}$ can be estimated by the conditions 
\begin{equation}   
N^{\rm exp}_\gamma(\omega_-=\frac{E_{eg}}{2}-2\frac{[m_0(1+\sigma_{m0})]^2}{E_{eg}})=N_{\rm CL} \;\;\; {\rm and}\;\;\;
{N^{\rm exp}_\gamma(\omega_+=\frac{E_{eg}}{2}-2\frac{[m_0(1-\sigma_{m0})]^2}{E_{eg}})=0 }
\label{eq:condm0}
\end{equation}
where $N_{\rm CL}$ is the minimum expected number of events
for which at least one event should be observed with a given Confidence Level  
in Poisson statistics.  For example, 
assuming that our naive experiment is  background free 
we set $N_{3\sigma}\simeq 5.9$.

\begin{figure}
\centering
\includegraphics[width=0.8\textwidth]{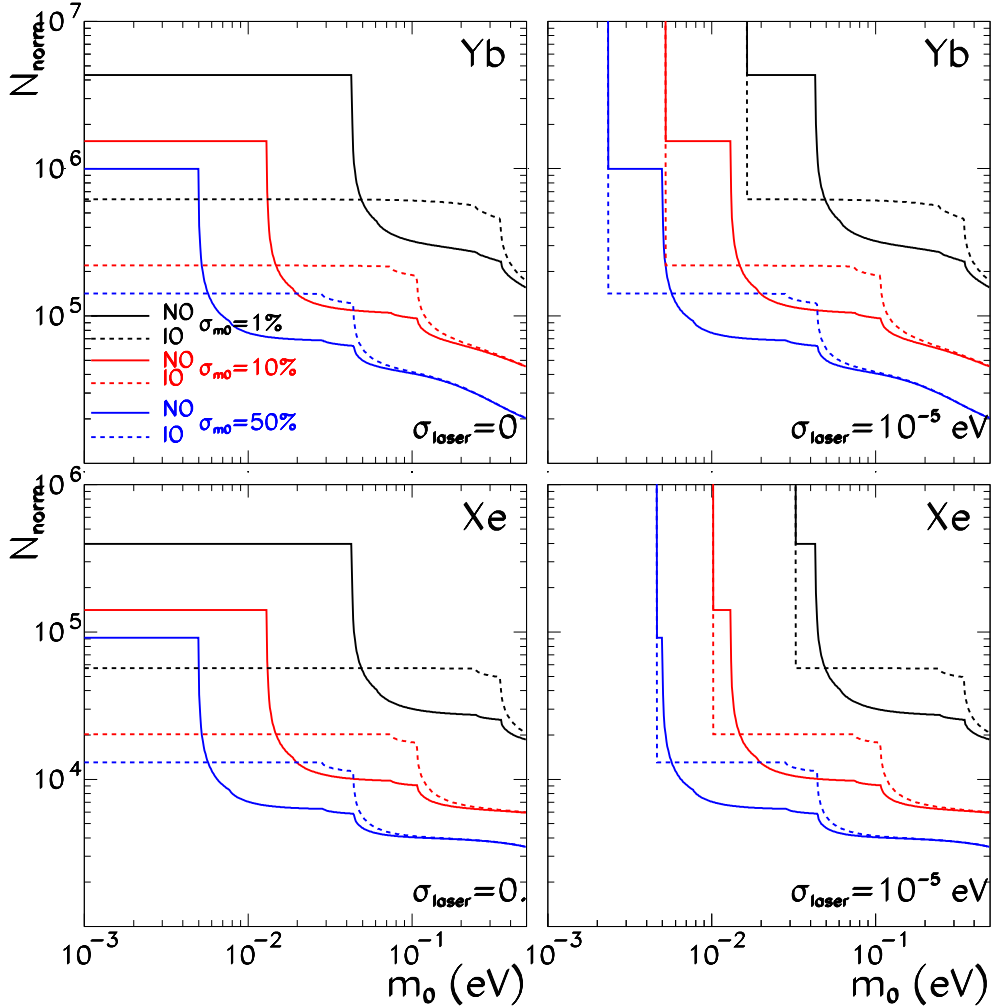}
\caption{
Required value of the rate normalization factor in Eq.(\ref{eq:Nnorm})
for the location of the end-point frequency of the RENP 
spectrum with 3$\sigma$ CL leading to a  precision in the
determination of the corresponding neutrino mass scale of 
$m_0\pm \sigma_{m_0}$ for three values of $\sigma_{m_0}=1,10,50$\% 
(black, red, and blue curves respectively) 
as a function of $m_0$. The full (dashed) lines correspond to NO (IO).
The upper (lower) panels are for Yb (Xe) atomic target. In the left  
panels infinite precision in the knowledge of laser frequency is assumed.
In the right panels the laser frequency is assumed to be known with 
$10^{-5}$  eV accuracy.} 
\label{fig:detm0}
\end{figure}

We plot in Fig.~\ref{fig:detm0} the required value of the normalization 
rate constant
\begin{equation}
N_{\rm norm}=
\left(\frac{T}{\rm s}\right) 
\left(\frac{V_{\rm tar}}{10^{2}\,{\rm cm}^{3}}\right) 
\left(\frac{n}{10^{21}\,{\rm cm}^{-3}}\right)^3 
\langle {\eta_{\w_0}} \rangle \; , 
\label{eq:Nnorm}
\end{equation}
to fulfill condition (\ref{eq:condm0}) as a function of 
$m_0$ and for different values of $\sigma_{m_0}$ . 
Notice that in writing Eq.~(\ref{eq:Nnorm})
we have neglected the $\omega$ dependence of the function 
$\langle\eta_\omega\rangle$ in the range $\omega_- \geq \omega\geq\omega_+$.
We show the results for an idealized case of 
perfect knowledge of the laser frequency and 
for laser with frequency known with  finite accuracy 
$\sigma_{\rm laser}=10^{-5}$ eV, which imposes the additional
constraint $\omega_+-\omega_-\leq \sigma_{\rm laser}$.

From the figure we see that if the accuracy at which the laser frequency
is known was infinite, the required normalization factor would always
be lower for Xe as a consequence of the 
larger decay  rate $\gamma_{pg}$, even though the level energies involved
are larger.  The inclusion of a finite accuracy for the
the laser frequency results in  cut-off values 
$m_{0,\rm min}$ below which the determination of $m_0$ is not possible. 
They  are given  by the condition 
$\omega_+-\omega_-\geq \sigma_{\rm laser}$ and, at a given $\sigma_{m_0}$, 
these maximum  reachable values are smaller for Yb than for Xe since  the
corresponding frequency differences are larger for Yb    
due to its smaller value of  $E_{eg}$. 
We also see that, the required normalization decreases as $m_0$ 
increases. This is so despite the overall normalization of 
$I(\omega)$  is lower for higher $m_0$ (see Fig.~\ref{fig:spec}).
But the larger is $m_0$ the larger is the difference between 
{$\w_+$ and $\w_-$},  so one is sampling the spectrum at 
lower values,  of the frequency, ie further from the final cutoff,
where $I(\omega)$ is relatively larger.

The horizontal asymptotes correspond to values of $m_0$ for which 
$\omega_-$ is above the previous to the last threshold, 
$\omega_->\omega_{12}$ {$(\omega_{31})$} for NO (IO), because the spectrum is
independent of $\omega$ in this range. 
The maximum value of $m_0$ for which this asymptotic constant rate 
normalization occurs is independent of the atomic target as it is purely set 
by the neutrino mass spectrum. 
It is reached at higher $m_0$ for IO than for NO since 
in NO the condition reads 
$2\,m_0(1+\sigma_{m_0})<(m_1+m_2)_{\rm NO}=
m_0+\sqrt{m_0^2+\Delta m^2_{21}}$  while for IO it is at
$2\,m_0(1+\sigma_{m_0})<(m_3+m_1)_{\rm IO}=m_0+\sqrt{m_0^2+ \Delta^2_{31}}$. 

\begin{figure}
\centering
\includegraphics[width=0.5\textwidth]{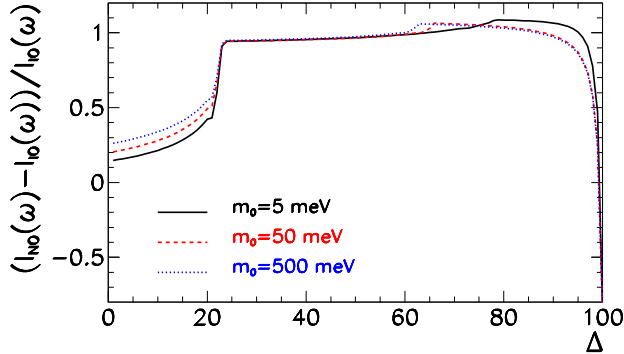}
\caption{Relative difference of the RENP spectra for
NO and IO as a function of the normalized frequency variable $\Delta$ 
defined in Eq.~(\ref{eq:omegaren}) for several values of 
$m_0$ as labeled in the figure.}
\label{fig:specnorm}
\end{figure}

\subsection{Discrimitation between Orderings}

Next we consider the minimum requirement for statistical discrimination
between the two orderings. We will assume that this is done after
the value of $m_0$ has been established. As seen in Fig.~\ref{fig:spec}
for a given value of $m_0$ the main difference between the two orderings
is the overall normalization  with the additional features associated
with the different location of the threshold frequencies. To illustrate
further the relative size of such features  we plot in 
Fig.~\ref{fig:specnorm} the relative difference between the NO and IO  
RENP spectra for Xe (the corresponding one for Yb is very similar)
plotted against a normalized frequency variable: 
\begin{eqnarray}
&&\Delta =20+80\frac{\omega-\omega^{\rm thres}_{\rm min}}
{\omega^{\rm thres}_{\rm max}-\omega^{\rm thres}_{\rm min}}\; ,  \label{eq:omegaren}\\
&& 
\omega^{\rm thres}_{\rm max}=\frac{E_{eg}}{2} - \frac{4m_0^2}{2E_{eg}}\; ,\\
&&\omega^{\rm thres}_{\rm min}=\frac{E_{eg}}{2} - \frac{4m_3^2}{2E_{eg}}\; , 
\;\;\; {\rm with}\;\;\; m_3^2=m_0^2+ \Delta m^2_{31}\;\;\; {\rm for\; NO}\;,\\
&&\omega^{\rm thres}_{\rm min}=\frac{E_{eg}}{2} - \frac{4m_2^2}{2E_{eg}}\;,
\;\;\;{\rm with}\;\;\; m_2^2=m_0^2- \Delta m^2_{32}\;\;\; {\rm for\; IO}\;.
\end{eqnarray}

As seen in the figure there are three main ``regions'' in the curves,
below the lowest threshold, in between the lowest and the 
previous-to-end point threshold, and above that previous-to-end 
point threshold. In view of this behaviour we foresee a naive  experiment 
which samples the spectra for three values
of the frequency, each one, corresponding to these three regions, so
we chose $\omega_{1,2,3}$ such that $\Delta_1=0$, $\Delta_2=40$, and 
$\Delta_3=80$. 
Using this information as input we study the requirements
for discrimination of the orderings following a similar approach to
Ref.~\cite{Blennow:2013oma}. 

\begin{figure}
\centering
\includegraphics[width=0.5\textwidth]{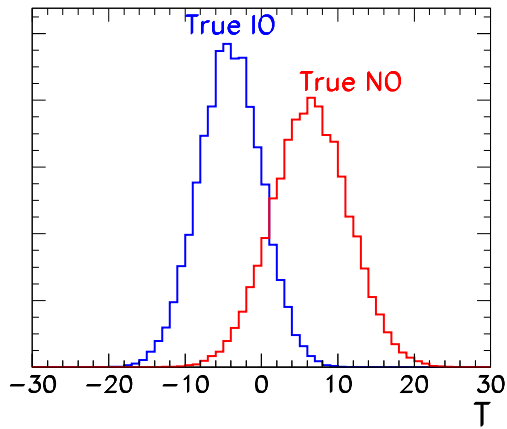}
\caption{Probability distribution for the $T$ test statistics 
in Eq.(\ref{eq:Tdef}) for events generated
about $N^{\rm obs}_{\gamma}(\omega_i)$ 
as expected for Xe with 
$\theta_{\rm true}$ corresponding to the best fit values and $m_0=0.01$ eV
and $N_{\rm norm}=3000$. 
The blue (red) histogram correspond to O$_{\rm true}$=IO (NO).}
\label{fig:Tdis}
\end{figure}

In brief, let's assume that the observed rates $N^{\rm obs}_{\gamma}(\omega_i)$
for $i=1,2,3$  are those expected for some values of the oscillation
parameters and some normalization rate $N_{\rm norm}$ 
for some ordering O$_{\rm true}$. 
Notice that for  simplicity we assume the true normalization to be the same 
for the three  frequencies.
We build the likelihood ${\cal L}$ function for that data to be described 
within a given ordering ``O''as 
\begin{equation}
\chi^2_{\rm RENP, O}=-2 \log({\cal L}^{\rm RENP}_{\rm O}]=2\, \begin{array}{c}{\rm min}\\[-0.1cm]
{\theta\subset {\rm O}}\end{array} 
\left[\,\sum_{i=1}^3
N^{\rm exp}_{\gamma}(\omega_i;\theta,{\rm O})
-N^{\rm obs}_{\gamma}(\omega_i)
-N^{\rm obs}_{\gamma}(\omega_i)
\log\left(\frac{N^{\rm exp}_{\gamma}(\omega_i;\theta,{\rm O})}
{N^{\rm obs}_{\gamma}(\omega_i)}\right)\right] \; ,
\end{equation}
where $N^{\rm exp}_{\gamma}(\omega_i;\theta,{\rm O})$ is the number of 
expected RENP events with frequency {$\w_i$} for parameters
$\theta$ 
(we label $\theta$ a given set of values for the  oscillation parameters  
and normalization)  and for the ordering ``O'' 
We then we  define the test statistics $T$ as
\begin{equation}
T=\chi^2_{\rm RENP, IO}-\chi^2_{\rm RENP, NO}\; .
\label{eq:Tdef}
\end{equation}
To determine the probability distribution of $T$ we generate pseudo experiments
Poisson distributed about $N^{\rm obs}_{\gamma}(\omega_i)$ 
and for each of them we compute the value of $T$. 
We show in Fig.\ref{fig:Tdis} as example the  distribution for the case 
in which  $N^{\rm obs}_{\gamma}(\omega_i)$ are those expected for Xe with 
$\theta_{\rm true}$ corresponding to the best fit values and $m_0=0.01$ eV
and $N_{\rm norm}=3000$. The blue (red) histogram corresponds 
O$_{\rm true}$=IO (NO), {\it i.e.} they are the distributions $p(T,{\rm IO})$
and $p(T,{\rm NO})$ respectively.
As expected $p(T,{\rm NO})$  is
peaked at positive values of $T$ (since in this case
$\chi^2_{\rm RENP, IO}$  is most likely larger than $ \chi^2_{\rm RENP, NO}$) while
the opposite holds for $p(T,{\rm IO})$. As $N_{\rm norm}$ increases the
distributions become more sharply peaked, so the overlap between them
decreases. 

The question we want to address is for what minimum value of 
$N_{\rm norm}$ the overlap is small enough so we can discriminate against 
the wrong ordering at a given CL, $1-\alpha$. In order to quantify this, 
we make use of the condition that the {\sl median sensitivity} is smaller 
than $\alpha$.
This condition imposes that the median of the  distribution with the 
right ordering  (ie the value of $T_c$ for which 50\% of the pseudo-experiments 
have  $T>T_c$ and 50\% have $T<T_c$)  has a probability smaller that 
$\alpha$, in the distribution of the wrong ordering. 
This is, for true NO we need to find
$N_{\rm norm}$  for which 
\begin{equation}
\int_{T_c^{\rm NO}}^\infty p(T,{\rm IO})\leq \alpha\; . 
\end{equation}
Conversely for true IO we need to find  $N_{\rm norm}$  for which 
\begin{equation}
\int^{T_c^{\rm IO}}_{-\infty} p(T,{\rm NO})\leq \alpha\; . 
\end{equation}

\begin{figure}
\centering
\includegraphics[width=0.8\textwidth]{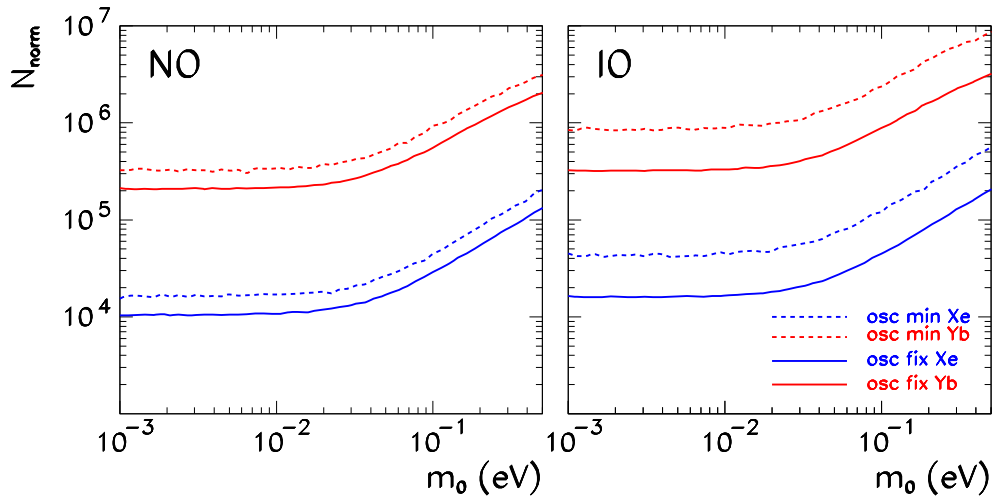}
\caption{
Required value of the rate normalization factor in Eq.(\ref{eq:Nnorm})
for which the median sensitivity is better than 99\% CL 
assuming that the true ordering is NO (left panel) and IO (right panel).
In each panel the two upper (lower) curves correspond to atomic target of
Yb (Xe). In the full lines  the oscillation parameters are kept fixed
to their best fit values given in Eq.~(\ref{eq:oscpar}). 
In the dashed lines they are minimized within the
present allowed ranges of the global oscillation analysis   
in Ref. \cite{Gonzalez-Garcia:2014bfa} (see text for details).}
\label{fig:detord}
\end{figure}

The result of this exercise is shown in Fig.~\ref{fig:detord}. 
In the figure we plot the minimum value of $N_{\rm norm}$ for which
the median sensitivity to discriminate between  orderings is 
99\% CL as a function of the neutrino mass scale $m_0$. 
In the left (right) panel the true ordering is NO (IO).
The full lines are obtained keeping the oscillation parameters
fixed to the best fit values of the present oscillation analysis
given in Eq.~(\ref{eq:oscpar}). The dashed lines include the effect
of the present uncertainty on the oscillation parameters.  
In doing so the oscillation parameters are minimized over  
within the present allowed ranges of the global oscillation analysis   
in Ref. \cite{Gonzalez-Garcia:2014bfa}. In order to include this
effect we add to $\chi^2_{\rm RENP}$ a gaussian bias for each of the
oscillation parameters  with central value and  1$\sigma$ error
given in Eq.~(\ref{eq:oscpar}). As seen in the figure, 
the inclusion of this uncertainty makes
the minimum required $N_{\rm norm}$  larger by a factor 
${\cal O}$(1.5--2.5). The results are also shown for the two atomic
targets considered, Xe (lower blue curves) and  Yb (higher red curves).
In the figure we also see that for $m_0\lesssim 0.03$ the result is independent
of $m_0$ while for heavier neutrino mass scales, the minimum $N_{\rm norm}$ 
required grows with $m_0$ because the sample values of $I(\omega_i)$ are 
lower as $m_0$ increases.  For the same reason the required $N_{\rm norm}$ 
is always larger for true IO than for true NO. 

\section{Summary and outlook }

In this work we have quantified the potential of 
macrocoherent atomic de-excitation via
radiative emission of neutrino pairs 
as a probe of the neutrino mass spectrum.
In particular we have evaluated the requirements for statistical 
determination of the most immediate unknowns of the neutrino spectrum: 
the neutrino mass scale and the mass ordering. 
In order to do so we have devised a minimum set of measurements 
and the associated statistical tests,  capable of determining those 
neutrino properties in an  idealized background free environment. We 
have considered two possible  atomic targets whose lowest  levels 
verify the conditions for RENP de-excitation: Xe and Yb. 

Our results are summarized in Figs.~\ref{fig:detm0} 
and ~\ref{fig:detord}.
Figure ~\ref{fig:detm0} displays the required value of the rate 
normalization factors  in Eq.(\ref{eq:Nnorm}) for the determination 
of the lightest neutrino mass $m_0$ with 3$\sigma$ CL and
given precision (1,10,50\%). Figure ~\ref{fig:detord} contains 
the corresponding results for the ordering determination at 99\% CL.

Our results are sobering. Regardless of the particular behaviour of the curves 
in the figures we find that even  under the ideal conditions assumed here,
the required normalization for the 3$\sigma$ determination of $m_0$ 
with accuracy better than 50\% implies experimental live times 
of the order of days to years for each frequency for a target of volume of 
order 100 cm$^3$ containing  about $10^{21}$ atoms per cubic centimeter
in a totally coherent state with maximum value of the electric field 
in the target ($\langle {\eta_{\w}}\rangle \sim {\cal O}(1)$).
Also comparing the results in Figs.\ref{fig:detm0} and \ref{fig:detord}
we find that in order to discriminate between the mass orderings 
at 99\% CL one needs similar target coherence and running time conditions 
to those required for the determination of the mass scale at $3\sigma$. 
Such conditions seem to be, as of today, way beyond the reach of our current 
technology.

\section*{Acknowledgments}
M.C.G-G and N.S are
supported by USA-NSF grant PHY-13-16617 and by FP7 ITN INVISIBLES
(Marie Curie Actions PITN-GA-2011-289442).  M.C.G-G also acknowledges
support by grants 2014-SGR-104 and by FPA2013-46570 and
consolider-ingenio 2010 program CSD-2008-0037.
J.J.G-C is also supported by the following agencies and
institutions: the European Research Council (ERC) under the
Advanced Grant 339787-NEXT; the Ministerio de Economia y Competitividad 
of Spain under grants CONSOLIDER-Ingenio 2010 CSD2008-0037 (CUP), 
FPA2009-13697-C04 and FIS2012-37947-C04.
\bibliography{biblio}

\end{document}